\documentclass[aps,pra,floatfix,nofootinbib,tightenlines,twocolumn,amsmath,amssymb]{revtex4}
\pdfoutput=1

\usepackage{amsmath,bbm,graphicx}
\usepackage[letterpaper=true,pdftex,pdfpagelabels=true,bookmarks = true,colorlinks = true, citecolor=blue, urlcolor=cyan]{hyperref}

\newcommand{\bra}[1]{\ensuremath{\left\langle #1\right\vert}}
\newcommand{\ket}[1]{\ensuremath{\left\vert #1\right\rangle}}
\newcommand{\expval}[1]{\ensuremath{\left\langle #1 \right\rangle}}
\newcommand{\hsp}[1]{\hspace{#1 em}}
\newcommand{\sqz}{\hsp{-0.1}}
\newcommand{\ketbra}[2]{\left\vert{#1}\right\rangle \sqz\sqz\sqz \left\langle{#2}\right\vert}
\newcommand{\braket}[2]{\left\langle{#1}\right\vert \sqz \sqz \left. {#2}\right\rangle}

\newcommand{\tr}{\text{Tr}}
\newcommand{\nbar}{\overline{n}}
\def\iden{\mathbbm{1}}

\graphicspath{{Figures/}}

\begin{document}
\title{Strong quantitative benchmarking of quantum optical devices}
\date{\today}

\author{N. Killoran and N. L\"{u}tkenhaus}
\affiliation{Institute for Quantum Computing and Department of Physics \&
Astronomy, University of Waterloo, N2L 3G1 Waterloo, Canada}

\begin{abstract}
Quantum communication devices, such as quantum repeaters, quantum memories, or quantum channels, are unavoidably exposed to imperfections. However, the presence of imperfections can be tolerated, as long as we can verify such devices retain their quantum advantages. Benchmarks based on witnessing entanglement have proven useful for verifying the true quantum nature of these devices. The next challenge is to characterize how strongly a device is within the quantum domain. We present a method, based on entanglement measures and rigorous state truncation, which allows us to characterize the degree of quantumness of optical devices. This method serves as a quantitative extension to a large class of previously-known quantum benchmarks, requiring no additional information beyond what is already used for the non-quantitative benchmarks. 
\end{abstract}

\maketitle

\section{Introduction}\label{sec:intro}

Quantum communication is the transmission of information via quantum states. Such states can exhibit correlations which are far stronger than anything allowed by classical theory. These quantum correlations enable us to perform novel communication tasks, such as the teleportation of quantum states \cite{bennett93a} or the distribution of secret cryptographic keys through public channels \cite{bennett84a}. To support the transfer and storage of quantum information, our communication infrastructure must incorporate new quantum devices such as quantum repeaters, quantum teleportation systems, quantum memories, and, most generally, quantum channels.  Unfortunately, unavoidable imperfections in these devices can degrade the embedded quantum information. Although we can deal with some degradation by employing classical postprocessing, if the imperfections become too strong we can lose the quantum advantage altogether. Thus, it is important to distinguish whether our communication devices are truly functioning in a quantum way.

\begin{figure}
	\includegraphics[width= 1.0 \columnwidth]{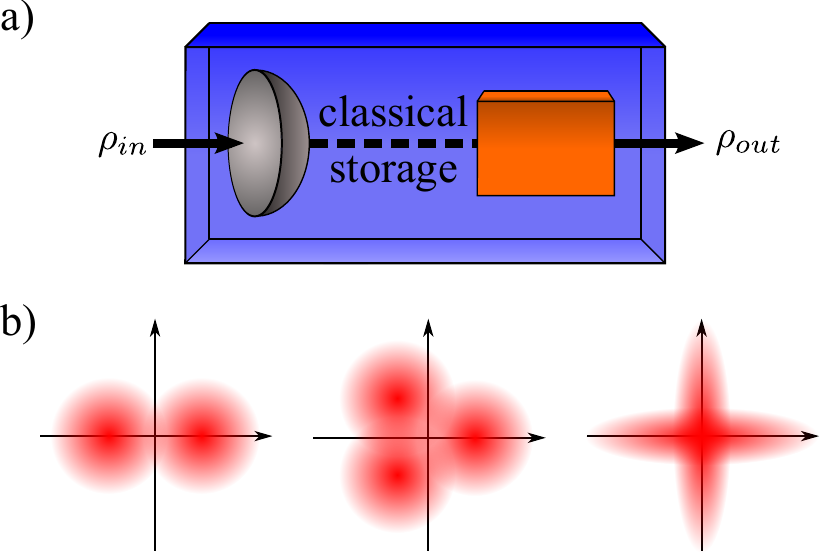}
	\caption{(Color online) a) Device implementing a measure and prepare strategy: A measurement is made on the input state, the result is stored or transmitted classically, and a new quantum state is prepared based on this measurement result.\\
	b) Example distributions of test states: Entanglement-based benchmarks have been designed for ensembles consisting of two or more coherent states on a ring, squeezed and antisqueezed vacuum, as well as for generalizations to mixed states.}
	\label{fig:benchmarking1}
\end{figure}

Given a quantum state as input, a classical device is one which uses a \emph{measure and prepare} strategy (see Fig. \ref{fig:benchmarking1}a). This strategy consists of a measurement on the state, storage or transmission of the measurement result classically, and preparation of a new output state based on this classical information. Any device which can be shown to be inconsistent with this strategy is said to operate in the \emph{quantum domain}. From this basic idea, we can develop benchmarks which tell us whether or not a device is in the quantum domain. For example, one approach is to use the fidelity as a figure of merit. If the fidelity between the output state and the input state, averaged over some ensemble of test states, is higher than the best fidelity achievable by a measure and prepare strategy, then the device is in the quantum domain. Unfortunately, this approach has some practical drawbacks. To calculate the fidelity accurately, we need to either perform (experimentally costly) full tomography on the output states or make unverified assumptions about them (e.g., that they are Gaussian). As well, the optimal classical fidelity must be found anew for each type of test ensemble, and it has only been found for a few cases \cite{braunstein00a, hammerer05a, namiki05a, namiki08a, calsamiglia09a, owari08a, adesso08a}. Moreover, many of these fidelity benchmarks require testing with an infinite ensemble of states, which is not possible in experiment.

Another approach to the quantum benchmarking problem relies on witnessing entanglement, since measure and prepare channels are mathematically equivalent to the so-called \emph{entanglement breaking channels} \cite{horodecki03a}. If we verify that a channel or device preserves entanglement, when acting on a subsystem, then we also know that the device operates in the quantum domain. In fact, this can even be done with a minimum of experimental resources. For one, this approach does not actually require the use of entangled states; instead, we can use `effective' entanglement \cite{bennett92c,curty04a,moroder06a,rigas06a}. As well, only a few test states and a few different measurements suffice to witness entanglement. This experimental simplicity, without the need for unverified assumptions, makes entanglement-based benchmarks very practical. This approach is especially useful for testing optical devices with continuous degrees of freedom \cite{lorenz06a,wittmann10a}, where the fidelity-based techniques cannot be easily used. Entanglement-based benchmarks have already been designed which use two \cite{rigas06a,haseler08a} or more \cite{haseler10a} coherent test states, two squeezed test states, and even mixed test states \cite{haseler09a}.

These benchmarks (fidelity- or entanglement-based) do not give the full story so far. Currently, they only address the simple question: \emph{quantum or classical?} There is a much richer realm of possibilities within the quantum domain. A device could be in the quantum domain, yet operate very inefficiently, consuming more resources than another device which performs the same task. To capture this difference in performance, we need quantitative extensions of existing benchmarks. Such measures will enable us to characterize and compare different devices within the quantum domain. For entanglement-based benchmarks, a natural extension is to consider \emph{how well} devices preserve entanglement, by making use of entanglement measures. Such an approach was recently studied in \cite{killoran10b}, calling this notion the \emph{quantum throughput}. This quantitative extension required no additional experimental resources beyond the (already limited) requirements of previous entanglement-based benchmarks. 

The results of \cite{killoran10b} are an important first step, but there remains much room for improvement. For one, the `quantitative domain' should be as faithful as possible with the quantum domain. In other words, if a device preserves the entanglement of some test state, then we should aim to give a non-trivial estimate of how much entanglement remains. We may not be able to estimate the amount exactly, but it is certainly non-zero. Unfortunately, the methods from \cite{killoran10b} only give non-trivial estimates for a small fraction of the quantum domain (see Sec. \ref{sec:entbench}), leading to overly stringent device requirements. Clearly there is a need for more work on the theoretical side. Additionally, \cite{killoran10b} considers only one type of benchmarking situation, involving two coherent test states. It is desirable to have quantitative extensions for many of the other entanglement-based benchmark schemes, and we need to develop new tools to handle these cases.

In this paper, we present a new method to quantitatively extend existing entanglement-based benchmarks for continuous variable optical devices. This new method is very successful, giving much stronger results than were previously achievable, often faithful over nearly all of the quantum domain. The method relies on projecting optical states to large finite subspaces and rigorously bounding the resulting truncation errors using homodyne measurement results. The strength of our approach can also be improved simply by devoting more computational resources to the task. The method is quite general and can be applied to many of the existing entanglement-based benchmarking situations, requiring no additional experimental resources. 

The remainder of this paper is organized as follows. In Sec. \ref{sec:entbench}, we outline the main notions of entanglement-based benchmarking, both qualitative and quantitative. Then, in Sec. \ref{sec:minent}, we present our new method for calculating quantitative benchmarks. We apply our scheme to a variety of scenarios and report on the results in Sec. \ref{sec:results}. We conclude the paper with Sec. \ref{sec:conclusion}, while detailing some proofs in Appendices \ref{app:firstmomentbounds} and \ref{app:secondmomentbounds}.

\section{Benchmarking quantum optical devices using entanglement}\label{sec:entbench}
\subsection{Qualitative benchmarking}\label{ssec:entbench}

Here, we briefly review some basic notions of entanglement-based benchmarking schemes. The interested reader is referred to the references for further details. These schemes are based around the fact that the set of measure and prepare channels is exactly the same as the set of entanglement breaking channels \cite{horodecki03a}. If we represent a device by a superoperator $\Lambda$, then it is entanglement breaking if the state $[\iden\otimes\Lambda](\rho)$ is separable \emph{for all $\rho$}. To disclude this behaviour, we must find a state which remains entangled after interaction with the device. To this end, consider the following entangled state:
\begin{equation}\label{eq:entstate}
	\ket{\psi^{ent}}_{AB} = \frac{1}{\sqrt{d}}\left[\sum_{k=0}^{d-1}\ket{k}_A\otimes\ket{\psi_k}_B\right],
\end{equation}
where Alice's subsystem is $d$-dimensional (using the computational basis) and Bob's subsystem is an optical mode. The $\ket{\psi_k}$ come from some fixed ensemble of nonorthgonal test states. The nonorthogonality of these states is encoded in Alice's reduced density matrix:
\begin{equation}\label{eq:rhoA}
	\bra{m}\rho_A\ket{n} = \frac{1}{d}\braket{\psi_m}{\psi_n}.
\end{equation}
Alice's subsystem is kept isolated, while Bob's subsystem is sent through the device. This leads to the final state
\begin{equation}\label{eq:rhodef}
	\rho_{AB} = \left[\iden\otimes \Lambda\right]\left(\ketbra{\psi^{ent}}{\psi^{ent}}\right).
\end{equation}
Since Alice's subsystem was kept isolated, the reduced density matrix of this final state is still given by Eq. (\ref{eq:rhoA}).

Now, Alice measures her subsystem using the projective measurement $\{\ketbra{i}{i}\}_{i=0}^{d-1}$. When Alice measures the state $\ket{j}$, Bob's subsystem is projected, in general, into an infinite-dimensional mixed state 
\begin{equation}
	\rho_B^j:=\Lambda(\ketbra{\psi_j}{\psi_j}).
\end{equation}
We call these `conditional states,' since they are conditioned on Alice's measurement outcome. Bob measures the conditional states using a balanced homodyne detection setup. This allows him to determine the expectation values $\expval{\hat{x}}$, $\expval{\hat{p}}$, and variances
\begin{align}
	\text{Var}(\hat{x})=&\expval{\hat{x}^2}-\expval{\hat{x}}^2, \\
	\text{Var}(\hat{p})=&\expval{\hat{p}^2}-\expval{\hat{p}}^2 
\end{align}
of the canonical quadratures operators,
\begin{equation}
	\hat{x} = \frac{1}{\sqrt{2}}(\hat{a}^\dagger+\hat{a}), ~\hat{p}=\displaystyle\frac{i}{\sqrt{2}}(\hat{a}^\dagger-\hat{a}).
\end{equation}
This procedure is repeated many times to get useful measurement statistics for each of the conditional states. The goal is for Alice and Bob to verify the continued presence of entanglement in the final state $\rho_{AB}$. The given information (homodyne measurement results and knowledge of $\rho_A$) turns out to be sufficient for this task.

\begin{figure}
	\includegraphics[width= 1.0 \columnwidth]{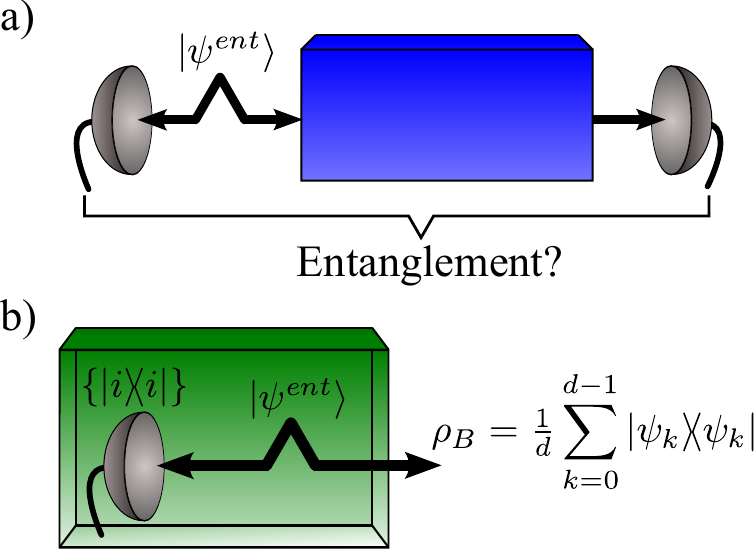}
	\caption{(Color online) a) Entanglement-based benchmarking: Is there any entanglement remaining after part of the state $\ket{\psi^{ent}}$ passes through the device? How much remains?\\
	b) `Effective' entanglement: Any source which produces a classical mixture of the test states $\ket{\psi_k}$ can be pictured as internally preparing the entangled state in Eq. (\ref{eq:entstate}) and performing a projective measurement on subsystem $A$.}
	\label{fig:benchmarking2}
\end{figure}

In fact, if we use a so-called `prepare and measure' scheme\footnote{Not to be confused with measure and prepare strategies.} \cite{bennett92c,rigas06a}, it is not necessary to actually use an entangled state to test if a device is entanglement breaking. Instead, we can use a classical mixture of the test states $\ket{\psi_k}$, and ask whether the device would have preserved entanglement if this mixture represented one subsystem of the entangled state Eq. (\ref{eq:entstate}). Any device which produces such a classical mixture can be thought of this way (see Fig. \ref{fig:benchmarking2}). Such entanglement is called \emph{effective entanglement}, and this way of thinking offers a great experimental advantage.

To map out the quantum domain, we need to paramaterize typical measurement results. For the quadratures, the action of any device on a test state can be described using the induced loss and noise. If the initial test states $\ket{\psi_k}$ have means $\expval{\hat{x}_{in}}_k$, $\expval{\hat{p}_{in}}_k$ and variances $\text{Var}_k(\hat{x}_{in})$, $\text{Var}_k(\hat{p}_{in})$, then the loss is parametrized by $1-T$, where
\begin{equation}
	\sqrt{T} = \frac{\expval{\hat{x}_{out}}_k}{\expval{\hat{x}_{in}}_k},
\end{equation}
and the excess noise is introduced through
\begin{equation}
	\text{Var}_k(\hat{x}_{out}) = \text{Var}_k(\hat{x}_{in}) + V_{ex}(x).
\end{equation}
Here, we assume that all conditional states exhibit the same loss and excess noise, and also that these quantities are symmetric with respect to $\hat{x}$ and $\hat{p}$. This is done merely for simplicity of presentation, and the entanglement verification or quantification methods do not rely on this choice. 

\begin{figure}
	\includegraphics[width= 1.0 \columnwidth]{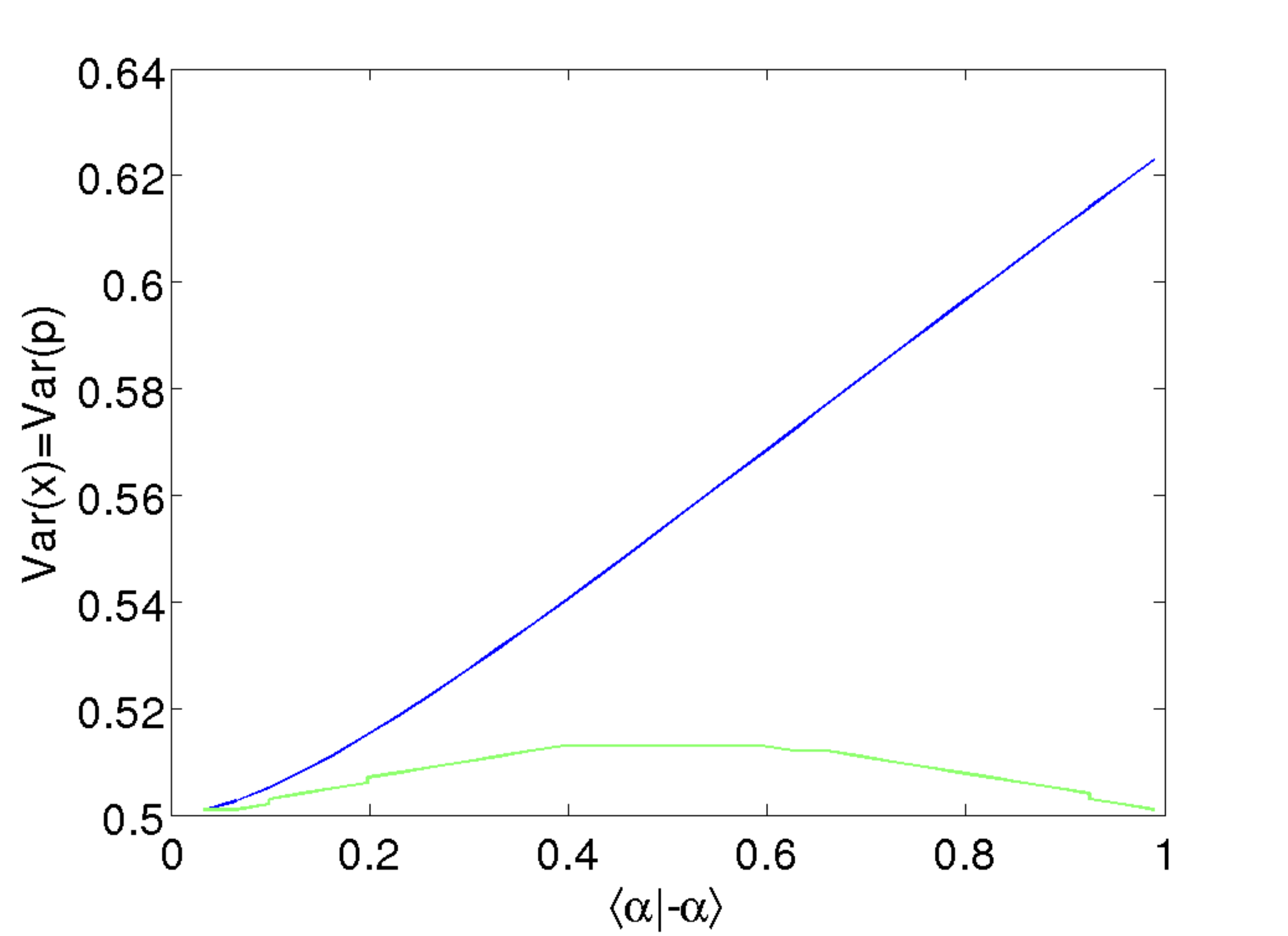}
	\caption{(Color online) Example benchmark results using two coherent states ($T=0.448$). The upper line was found using the EVM method; any data below this line is in the quantum domain. The lower line represents the limits of the quantitative domain using the methods of \cite{killoran10b}. There is clearly room for improvement on the quantitative benchmarks.}
	\label{fig:qdomaincompare}
\end{figure}

One tool which has proved quite useful for the task of entanglement verification is the so-called expectation value matrix (EVM) \cite{rigas06a}, which has been successfully used in all of the benchmarking scenarios mentioned above \cite{haseler08a,haseler10a,haseler09a} (with suitable substitutions in Eq. (\ref{eq:entstate})). The EVM method allows us to determine, for a given set of test states, which measurement results provide evidence that a device operates the quantum domain. An example result of the EVM method is shown in Fig. \ref{fig:qdomaincompare}, for the case of testing with two coherent states $\ket{\psi_k}\in\{\ket{\alpha},\ket{-\alpha}\}$.

\subsection{Quantitative benchmarking}\label{ssec:entbenchquant}

The main focus of this paper is in quantitative extensions to the above device testing scenario. The overall goal is to quantify how well an optical device preserves quantum correlations (the quantum throughput) by estimating the amount of entanglement, with respect to some measure $\mathcal{E}$, that remains in the output state $\rho_{AB}$ of Eq. (\ref{eq:rhodef}). We will develop the entanglement quantification method in general, but later use the negativity \cite{zyczkowski98a,vidal02a,lee00a} in concrete examples. The available information is the same as what is used for the qualitative benchmarking, namely homodyne measurements on the conditional states $\rho_B^j$ and knowledge of $\rho_A$. Since this information is tomographically incomplete, the best we can do is to give a lower bound on the entanglement, by minimizing over all states $\sigma_{AB}$ which have the same expectation values \cite{plenio06a}:
\begin{align}\label{eq:minent}
	\mathcal{E}_{min}(\rho_{AB}) = 
	\begin{cases}
	    	\displaystyle\min_{\sigma_{AB}\geq 0} &\mathcal{E}(\sigma_{AB})\\
	  s.t.~ &\tr(\sigma_{AB}\ketbra{i}{i}\otimes\hat{M_k})=\tr(\rho_B^i\hat{M_k})\\
	  &\tr(\sigma_{AB}\ketbra{i}{j}\otimes\iden_B)=\bra{j}\rho_A\ket{i}
	 \end{cases}.
\end{align}
The measurement operators $\hat{M}_k$ may be something directly measured in experiment (such as the quadratures) or something more abstract (such as projections onto pure states).

First results for this quantification problem were presented in \cite{killoran10b}. The underlying challenge is that the states in question are, in general, infinite-dimensional, making quantification inherently difficult. The basic strategy introduced to tackle this was to consider a projection of the infinite-dimensional mode subsystem down to a more manageable finite-dimensional subspace. This is more than just a practical simplification. It has been shown that, for any state $\rho_{AB}$, the presence of entanglement can be verified by searching within some finite subspace \cite{sperling09a}. In fact, for any entanglement measure with the strong monotonicity property, the entanglement content of the projected state gives a lower bound to the entanglement of the true state $\rho_{AB}$. 

To see this, let $P$ be a projector onto the desired mode subspace, and denote the (unnormalized) projected state by
\begin{equation}
	\rho_P := (\iden_A\otimes P)\rho_{AB}(\iden_A\otimes P).
\end{equation}
Then we have (with $Q:=\iden_B-P$ and $p=\tr(\rho_P)$)
\begin{align}\label{eq:projectionbound}
	\mathcal{E}(\rho_{AB}) \geq & ~p\mathcal{E}\left(\frac{(\iden_A\otimes P)\rho_{AB}(\iden_A\otimes P)}{p}\right)\notag\\
	 & +(1-p)\mathcal{E}\left(\frac{(\iden_A\otimes Q)\rho_{AB}(\iden_A\otimes Q)}{1-p}\right)\notag\\
	\geq &~p \mathcal{E}\left(\frac{\rho_P}{p}\right).
\end{align}
For practical purposes, it is convenient to absorb the prefactor $p$ and work directly with the unnormalized states ($0\leq \tr\rho_P\leq 1$). As such, we restrict to entanglement measures which are well-defined for unnormalized states and where, for any $p\in[0,1]$ and $\tau\geq0$,
\begin{equation}
	p\mathcal{E}(\tau)\geq \mathcal{E}(p\tau).
\end{equation}
In this case, we end up with the final inequality
\begin{equation}
	\mathcal{E}(\rho_{AB}) \geq \mathcal{E}(\rho_P).
\end{equation}
The above constraints must be kept in mind when choosing an entanglement measure.

To go along with this projection, we must also relax the measurement constraints in Eq. (\ref{eq:minent}). Importantly, even if we know an expectation value of $\rho_{AB}$ exactly, we may not be able to determine the corresponding expectation value of $\rho_P$ exactly. However, we should try to constrain the expectation values of the projected state to be as close as possible to those of the state. If we understand the system well enough, we can bound how much the expectation values can vary, ideally restricting them to some convex set $\mathcal{C}$ which depends on the known measurement results. If we can find such constraints, then we can perform a minimization over all projected states $\sigma_P$ compatible with the known constraints about the actual $\rho_P$ (and hence about $\rho_{AB})$:
\begin{align}\label{eq:minentproj}
	\mathcal{E}_{min}(\rho_{P}) = 
	\begin{cases}
	    	\displaystyle\min_{\sigma_{P}\geq 0} &\mathcal{E}(\sigma_{P})\\
	  s.t.~ &\tr(\sigma_{P}\ketbra{i}{i}\otimes\hat{M_k})\in\mathcal{C}(\rho_B^i,\hat{M}_k)\\
	  &\tr(\sigma_{P}\ketbra{i}{j}\otimes\iden_B)\in\mathcal{C}(\rho_A)
	 \end{cases}.
\end{align}

Finally, by solving the optimization of Eq. (\ref{eq:minentproj}), we will arrive at a rigorous lower bound on the entanglement of the infinite-dimensional output state $\rho_{AB}$,
\begin{equation}
	\mathcal{E}(\rho_{AB})\geq\mathcal{E}_{min}(\rho_P).
\end{equation}
To summarize, instead of finding the minimal entanglement over infinite-dimensional states with exactly known expectation values, we find the minimal entanglement over finite states with inexact expectation values. This latter minimization gives a lower bound to the former. Any choice of projection requires us to find appropriate constraints on the expectation values, and the tighter the constraints we can find, the stronger the results will be.

But what subspace should be projected onto? In \cite{killoran10b}, which considered only the testing scenario involving two coherent states ($\ket{\psi_i}\in\{\ket{\alpha},\ket{-\alpha}\}$), the choice is made to project onto a two-qubit subspace, the smallest subspace which can show entanglement. In this case, the operators $\hat{M}_k$ are projections onto the `largest' eigenvectors of the final mode states $\rho_B^i$ $(i=0,1)$. Constraints on these eigenvectors can be determined from the homodyne measurements using some bounds from the literature \cite{rigas06b,zhao09a}. A typical result of this approach is presented in Fig. \ref{fig:qdomaincompare}. The quantum domain, found using the EVM method, is shown in the same figure. 

Although the two qubit projection works, it is clear that there are some drawbacks, since the quantitative domain does not come close to matching the quantum domain. First, by projecting into the smallest possible subspace, we may be cutting out important information. Intuitively, we expect most information about our states to be confined to a small subspace, but a two-qubit subspace appears to be too small. Second, the employed bounds are only tight for very low noise values; this weakens the minimization at higher noise levels. Finally, the bounds are applicable only to a specific two-qubit subspace, and not readily generalized to higher dimensions.

In the following section, we propose a modified version of this quantification scheme which proves to be much more successful. We still make use of the projection to a finite subspace, but we choose a different type of subspace which can have significantly larger dimension. For this, we will derive new, rigorous bounds which allow us to make much tighter approximations on our state. This new method does not require any additional information, and unlike the previous quantification scheme, is readily generalized to other benchmarking scenarios. 

\section{Projecting onto Fock states}\label{sec:minent}

\subsection{Rigorous bounds on the projection errors}\label{ssec:bounds}

Perhaps the most natural choice of projection in the infinite-dimensional Hilbert space of an optical mode is a projection onto the energy eigenstates, namely the Fock states $\{\ket{n}\}_{n=0}^{N}$. Intuitively, the larger the cutoff $N$ is, the closer the projected state $\rho_P$ should be to the full state $\rho_{AB}$, and for $N\rightarrow\infty$, the two states are the same. Although we will use the finite projection trick of \cite{killoran10b} to find entanglement bounds, the projection onto Fock states requires a much different set of tools for its analysis than the previously considered two-qubit projection. 

Most importantly, we make the observation that it is not necessary to send $N$ to infinity. As long as the cutoff dimension $N$ is large compared to the mean photon number $\nbar$, a state and its projected form will be nearly identical. Why is this so? Because high Fock levels contribute more to the state's energy than low Fock levels. If we know $\nbar$, we can put strong bounds on the contributions from high Fock levels. We make this intuition concrete with the following lemma.

\emph{Fock cutoff lemma}. Let $\tau$ be the state of a single optical mode, with mean photon number $\tr(\tau\hat{n})=\nbar<\infty$. Let $\tau_N$ be the (unnormalized) projection of $\tau$ up to the Fock level $N>0$, 
\begin{equation}
	\tau_N = \sum_{k=0}^{N}\bra{k}\tau\ket{k}\ketbra{k}{k},
\end{equation}
and denote the expectation value of $\tau_N$ with respect to $\hat{n}$ by
\begin{equation}
	\tr(\tau_N\hat{n})=:\nbar_N.
\end{equation}
Then the following inequality holds:
\begin{equation}\label{eq:focklemmabound}
	\tr(\tau)-\tr(\tau_N)\leq \frac{\nbar-\nbar_N}{N+1}.
\end{equation}
\emph{Proof:} For $k>0$, parametrize the diagonal entries of $\tau$ by
\begin{equation}
	\tau_k = \bra{k}\tau\ket{k} =: \frac{c_k}{k}.
\end{equation}
The coefficients $c_k$ are positive, but otherwise unknown. We have
\begin{equation}
	\tr(\tau\hat{n})=\sum_{k=0}^\infty k\tau_k = \sum_{k=1}^\infty c_k=\nbar, 
\end{equation}
and
\begin{equation}
	\tr(\tau_N\hat{n})= \sum_{k=1}^N c_k=\nbar_N. 
\end{equation}
This leads to 
\begin{align}
	\tr(\tau)-\tr(\tau_N)  = & \displaystyle\sum_{k=N+1}^{\infty}\frac{c_k}{k}\\
	 \leq & \displaystyle\sum_{k=N+1}^{\infty}\frac{c_k}{N+1}\\
	 = & \frac{1}{N+1}\left(\sum_{k=1}^{\infty}c_k-\sum_{k=1}^{N}c_k\right)\\
	 = & \frac{\nbar-\nbar_N}{N+1}.
\end{align}
Finally, note that this upper bound can actually be saturated, for example by the state
\begin{equation}
	\tau = \left(1-\frac{\nbar}{N+1}\right)\ketbra{0}{0}+\frac{\nbar}{N+1}\ketbra{N+1}{N+1}.
\end{equation}
\hfill$\square$

This lemma provides an upper bound on how different the trace of a state is from the trace of its projected form. Since the mean photon number can be inferred from homodyne measurements, 
\begin{equation}
	\nbar := \tr(\tau\hat{n}) = \tr\left(\tau\frac{\hat{x}^2+\hat{p}^2-\iden}{2}\right),
\end{equation}
we can easily apply this lemma to the present benchmarking scenario. In this case, the quantity $\nbar_N$ will be a free parameter which is a linear combination of diagonal elements from $\tau_N$. It is only constrained by the obvious bounds $0\leq\nbar_N\leq\nbar$. Although it would be valid to replace the right hand side of Eq. (\ref{eq:focklemmabound}) with the strictly numeric quantity $\nbar/(N+1)$, it is more useful to use the bound as is, since it links the constraints on $\tr(\tau_N)$ and $\nbar_N$. Most importantly, since $\nbar$ is fixed, the bound can be made arbitrarily tight by increasing the cutoff $N$. Along the same lines,  we can derive similar constraints (also depending on $\nbar$ and $N$) for all the other important expectation values. Together, these constraints will form the basis of our new entanglement quantification method, outlined in the next subsection. 

For instance, if a state $\tau$ has some known values of $\expval{\hat{x}}$ and $\expval{\hat{p}}$, then the expectation values of the same operators with respect to the truncated state $\tau_N$ cannot be entirely arbitrary. In fact, if $N$ is large enough, the truncated state's quadratures have to be quite close to the infinite-dimensional state's quadratures. We can make this argument explicit with the following bound, involving the ladder operator $\hat{a}=\frac{1}{\sqrt{2}}(\hat{x}+i\hat{p})$: 
\begin{equation}
	|\tr(\tau\hat{a})-\tr(\tau_N\hat{a})| \leq \epsilon,\label{eq:abound}
\end{equation}
with
\begin{equation}\label{eq:epsilon}
	\epsilon:=\sqrt{(\nbar-\nbar_N)(\tr(\tau)-\tr(\tau_{N-1}))},
\end{equation}
where $\tau_{N-1}$ is the truncation of the state $\tau$ up to Fock level $N-1$. This constraint can be straightforwardly derived by expanding the annihilation operator in the Fock basis and applying the triangle and Cauchy-Schwarz inequalities, as well as appealing to the positivity of $\tau$. The full derivation can be found in Appendix \ref{app:firstmomentbounds}. Although the quantity $\tr(\tau\hat{a})$ is not directly measureable, it can be found from the expectation values of the quadratures $\hat{x}$ and $\hat{p}$ with respect to $\tau$. As well, the bound $\epsilon$ is not fixed, since it depends on other parameters. However, by the Fock cutoff lemma, it has a maximum size
\begin{equation}
	\epsilon\leq\frac{\nbar}{\sqrt{N}}.
\end{equation}
Thus, it is clear that the bound ($\ref{eq:abound}$) can also be made tighter by increasing $N$. In practice, we will use the stricter adaptive constraint, Eq. (\ref{eq:epsilon}).

The other important operators for our problem are $\hat{x}^2$ and $\hat{p}^2$. Instead of directly working with these, it is useful to consider their difference, 
\begin{equation}
	\hat{d}:=\hat{x}^2-\hat{p}^2=\hat{a}^{\dagger 2}+\hat{a}^2.
\end{equation}
Although this operator is itself an observable, the required expectation values can be readily found from the available homodyne data. The original operators can be recovered using the relations
\begin{align}
	\hat{x}^2=&\frac{1}{2}\left[ \hat{d} + 2\hat{n}+\iden \right],\\
	\hat{p}^2=&\frac{1}{2}\left[ -\hat{d} + 2\hat{n}+\iden \right].
\end{align}
In a similar manner as before, we can bound the difference between $\expval{\hat{d}}$ for the measured infinite-dimensional state and for its truncation:
\begin{equation}\label{eq:differencedifference}
	\left| \tr(\tau\hat{d})- \tr(\tau_N\hat{d})\right| \leq \delta,
\end{equation}
with
\begin{equation}\label{eq:delta}
	\delta:=2\sqrt{\left(\nbar-\nbar_N\right)\left[ (\nbar-\nbar_{N-2})+(\tr(\tau)-\tr(\tau_{N-2})) \right]}.
\end{equation}
The proof of this bound uses similar arguments as the $\epsilon$ bound (see Appendix \ref{app:secondmomentbounds} for details). 

The final source of information in our benchmarking scheme is the reduced density matrix $\rho_A$, which enters in expectation values of the form
\begin{equation}\label{eq:offdiagexpval}
  \tr(\sigma_{AB}\ketbra{i}{j}\otimes\iden_B)=\bra{j}\rho_A\ket{i}.
\end{equation}
To see how to include this type of information, let $\tau_{AB}$ be a bipartite state, with reduced density matrix $\tau_A=\tr_B(\tau_{AB})$. As above, let $P$ be a projector on the B subsystem up to Fock level $N$, and let $Q=\iden_B-P$ (which implies $QP=PQ=0$). We also let
\begin{equation}
	\tau_N := (\iden_A\otimes P)\tau_{AB}(\iden_A\otimes P),
\end{equation}
and similarly define $\tau_Q$ using $Q$. Using the given relations, we find
\begin{align}
	\tau_A = & \tr_B(\tau_{AB}(\iden_A\otimes[P+Q]^2))\\
	 = &\tr_B(\tau_{AB}(\iden_A\otimes P)^2)+\tr_B(\tau_{AB}(\iden_A\otimes Q)^2).
\end{align}
But 
\begin{equation}
	\tr_B(\tau_{AB}(\iden_A\otimes P)^2)=\tr_B((\iden_A\otimes P)\tau_{AB}(\iden_A\otimes P)),
\end{equation}
(and similarly for the $Q$ term), so that
\begin{equation}
	\tau_A = \tr_B(\tau_N)+\tr_B(\tau_Q).
\end{equation}
Since $\tau_N$ and $\tau_Q$ are positive matrices, so are their reduced forms $\tau_{NA}:=\tr_B(\tau_N)$ and $\tau_{QA}:=\tr_B(\tau_Q)$. Dropping $\tau_{QA}$, we are left with
\begin{equation}\label{eq:rhoAbound}
	\tau_A\geq\tau_{NA}.
\end{equation}

Although it may not be obvious at this point, Eq. (\ref{eq:rhoAbound}) can be used to tightly constrain expectation values like Eq. (\ref{eq:offdiagexpval}) for the projected state. As with the other bounds, increasing the cutoff $N$ will strengthen the constraints. We will elaborate further about how to use this inequality below. The constraints found in Eqs. (\ref{eq:abound}-\ref{eq:epsilon}), (\ref{eq:differencedifference}-\ref{eq:delta}), and (\ref{eq:rhoAbound}) form the basis for determining a rigorous lower bound to a state's entanglement content. We explore how to do this in practice in the next subsection.

\subsection{Finding the minimal entanglement with semidefinite programming}
These new constraints can be directly applied to the problem of quantitative benchmarking. We know the expectation values of $\iden,\hat{a},$ and $\hat{d}$ for the conditional states $\rho_B^i$ of $\rho_{AB}$. These serve as the operators $\hat{M_k}$ found in the optimizations (\ref{eq:minent}) and (\ref{eq:minentproj}). As well, we have the entries of the reduced density matrix $\rho_A$. Using the above constraints, we can restrict these same expectation values for the projected state $\rho_N$ (we will use the subscript $N$ instead of $P$ from now on to emphasize the Fock state projection). We search through the projected subspace for the state with the minimal entanglement. The above bounds constrain more strictly where we can search, forming the convex regions $\mathcal{C}(\rho_B^i,\hat{M_k})$ and $\mathcal{C}(\rho_A)$ in the finite-dimensional optimization (\ref{eq:minentproj}). The larger we make the cutoff $N$, the tighter the constraints will be, and asymptotically, these bounds become equality constraints, making the optimizations (\ref{eq:minent}) and (\ref{eq:minentproj}) the same. Finally, we also enforce the obvious positivity ($\sigma_N\geq 0$) and trace ($\tr(\sigma_N)\leq 1$) constraints. 

To solve the finite-dimensional optimization (\ref{eq:minentproj}), we turn to semidefinite programming. In principle, the  matrices representing the operators $\hat{x}$ and $\hat{p}$ in the Fock basis are infinite-dimensional. This can lead to problems when working with them numerically. However, for a state which only has support up to Fock level $N$, there is no cause for concern. Such expectation values can be implemented by using truncated versions of the quadrature operators themselves. Indeed, if $\hat{x}_N$ is the (bounded) projection of $\hat{x}$ up to Fock level $N$, then 
\begin{equation}
	\tr(\tau_N\hat{x})=\tr(\tau_N\hat{x}_N).
\end{equation}
Expectation values of the quadratures are simply finite linear combinations of elements from $\tau_N$. The same idea holds as well for higher powers of the quadratures, as in the operator $\hat{d}$.

Conveniently, the above constraints can all be written in the form of matrix inequalities. For instance, for any $\tau$ which is positive, the constraint in Eqs. (\ref{eq:abound}-\ref{eq:epsilon}) may be recast as
\begin{equation}
	\begin{bmatrix}
		\nbar-\nbar_N & \tr(\tau\hat{a})-\tr(\tau_{N}\hat{a})\\
		\tr(\tau\hat{a}^\dagger)-\tr(\tau_{N}\hat{a}^\dagger) & \tr(\tau)-\tr(\tau_{N-1})
	\end{bmatrix} \geq 0.
\end{equation}
Similarly, we can rewrite the second order constraint, Eqs. (\ref{eq:differencedifference}-\ref{eq:delta}), as the constraint
\begin{equation}
		\begin{bmatrix}
		4(\nbar-\nbar_N) & \tr(\tau\hat{d})-\tr(\tau_{N}\hat{d})\\
		\tr(\tau\hat{d})-\tr(\tau_{N}\hat{d}) & \nbar-\nbar_{N-2} + \tr(\tau)-\tr(\tau_{N-2})
	\end{bmatrix} \geq 0.
\end{equation}
The original constraints follow from considering the positivity of the determinants. These matrix inequality constraints can be applied for each conditional state $\rho_B^i$.

Finally, we have the constraint 
\begin{equation}\label{eq:rhoNA}
	\rho_A\geq\rho_{NA}. 
\end{equation}
To see how this applies in a benchmarking situation, we note that the output state $\rho_{AB}$ can be written in the block form
\begin{equation}
	\rho_{AB} = \frac{1}{d}\begin{bmatrix}
		\rho_{00} & \rho_{01} & \rho_{02} &\cdots\\
		\rho_{10} & \rho_{11} & \rho_{12} &\\
		\rho_{20} & \rho_{21} & \rho_{22} &\\
		\vdots & & &\ddots
	\end{bmatrix},
\end{equation}
where the blocks $\rho_{ii}$ correspond to the conditional states $\rho_B^i$. The $(i,j)$ element of $\rho_A$ is exactly equal to $\tr(\rho_{ij})/d$. We can similarly decompose $\rho_N$ into a block form where the matrices $\rho_{ij}$ are replaced by their projected forms $\rho_{ij}^N$ up to Fock level $N$. The $(i,j)$ element of $\rho_{NA}$ is analogously equal to $\tr(\rho_{ij}^N)/d$. The inequality (\ref{eq:rhoNA}) allows us to constrain the trace of each block $\rho_{ij}^N$. To see this, consider a scenario where subsystem A is two dimensional. Then matrix inequality (\ref{eq:rhoNA}) becomes
\begin{equation}
		\begin{bmatrix}
		\tr(\rho_{00})-\tr(\rho_{00}^N) & \tr(\rho_{01})-\tr(\rho_{01}^N)\\
		\tr(\rho_{10})-\tr(\rho_{10}^N) & \tr(\rho_{11})-\tr(\rho_{11}^N)
	\end{bmatrix} \geq 0.
\end{equation}
Taking the determinant, we get 
\begin{equation}
	|\tr(\rho_{01})-\tr(\rho_{01}^N)|\leq\gamma,
\end{equation}
where
\begin{equation}
	\gamma:=\sqrt{(\tr(\rho_{00})-\tr(\rho_{00}^N))(\tr(\rho_{11})-\tr(\rho_{11}^N))}.
\end{equation}
This constraint is therefore very similar to the other ones, and it should be clear that we can make it stronger by increasing the cutoff $N$. We should point out that if subsystem A has more than two dimensions, the inequality (\ref{eq:rhoNA}) becomes even stronger since it also forces higher order determinants to be positive. 

We now have the full set of constraints on $\rho_N$ for the minimization, but we have yet to fix our objective function, i.e. the actual entanglement measure. As mentioned earlier, in order for the finite projection to work as desired, the measure should have the strong monotonicity property and also allow us to absorb prefactors and work with unnormalized states. As well, the measure (or a lower bound on it) should be computable within the framework of semidefinite programming. To this end, we choose the negativity \cite{zyczkowski98a,vidal02a,lee00a}, though this is not the only possible choice. Although the negativity may not have as nice an operational interpretation as some other entanglement measures, its main strength is that it is efficiently computable. For a Hermitian matrix, the negativity can be written as the following semidefinite program\footnote{For comparison purposes, it is important to note that that in \cite{killoran10b} the negativity is rescaled by a factor of $2$ from the formula used here.}:
\begin{align}
	\mathcal{N}(\tau) =  &\min \tr~\tau_-\\
	 & s.t.~\tau^{T_A} = \tau_+-\tau_-,\notag\\
	 & \tau_+\geq 0,~ \tau_-\geq 0,\notag
\end{align}
where $\tau^{T_A}$ is the partial transpose of $\tau$.

\section{Results}\label{sec:results}
The new quantification method can be used to extend many existing entanglement-based benchmarks. We will present results for the following ensembles of test states: a) two coherent states \cite{rigas06b,haseler08a}, b) two squeezed states \cite{haseler09a}, and c) three coherent states \cite{haseler10a}. All optimizations were done in Matlab with the solver SDPT3 \cite{toh99a} and the frontend YALMIP \cite{lofberg04a}, using a Dell Optiplex 760 desktop PC with a 3 GHz dual-core processor and 4 GB of RAM.

\begin{figure}
	\includegraphics[width= 1.0 \columnwidth]{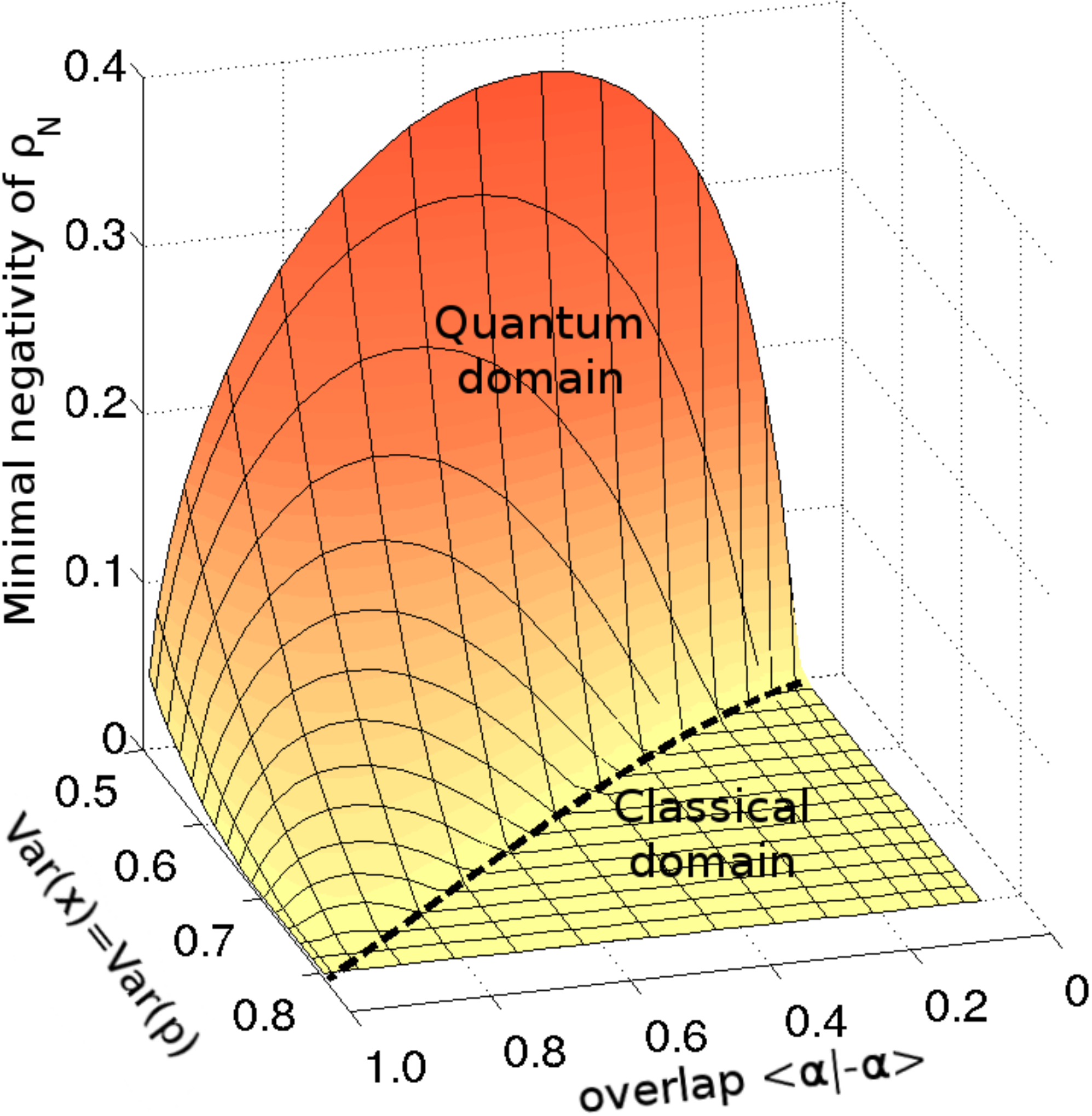}
	\caption{(Color online) Entanglement bounds for the two coherent state benchmark, using our new quantification scheme (with $N=20$), assuming no loss ($T=1.0$). The dashed black line represents the boundary of the quantum domain, found using the EVM method. Variances lower than this line correspond to devices within the quantum domain, whereas variances higher than this line are classical \cite{haseler08a} (note that the axes are inverted in order to see the results better). Direct comparison with the results of \cite{killoran10b} (Fig. 2a) shows a dramatic improvement in the entanglement bounds at higher variances.}
	\label{fig:twocoh}
\end{figure}

First, our results for the two coherent state benchmark are shown in Fig. \ref{fig:twocoh}, where we have used a cutoff at $N=20$ and fixed the loss parameter as $T=1.0$. We can see that the results are very faithful, being non-zero throughout nearly all of quantum domain. For other values of $T$, where the quantum domain is smaller, the results are similarly faithful. Only in the extreme regions of low overlap/low variance and high overlap/high variance are there small differences between the quantitative domain and the quantum domain. Both of these cases correspond to high values of $\nbar$, where the constraints are the weakest. But we expect very little effective entanglement can be found in these regions anyway. The region where $\braket{\alpha}{-\alpha}\rightarrow 1$ has very little initial entanglement (since the states are nearly identical), and there will be even less after subsystem B passes through the device. On the other hand, in the region where $\braket{\alpha}{-\alpha}\rightarrow 0$, $\rho_A$ becomes diagonal and our source replacement trick is useless. Although there may be a lot of real entanglement, we cannot verify any effective entanglement with a diagonal $\rho_A$. In principle, the quantitative domain could be extended by raising the cutoff $N$. Since in the limit $N\rightarrow\infty$, the optimization problem (\ref{eq:minentproj}) converges to (\ref{eq:minent}), we conjecture that any mismatch between the quantitative domain and the quantum domain can be resolved by using sufficiently large computational resources.

\begin{figure}
	\includegraphics[width= 1.0 \columnwidth]{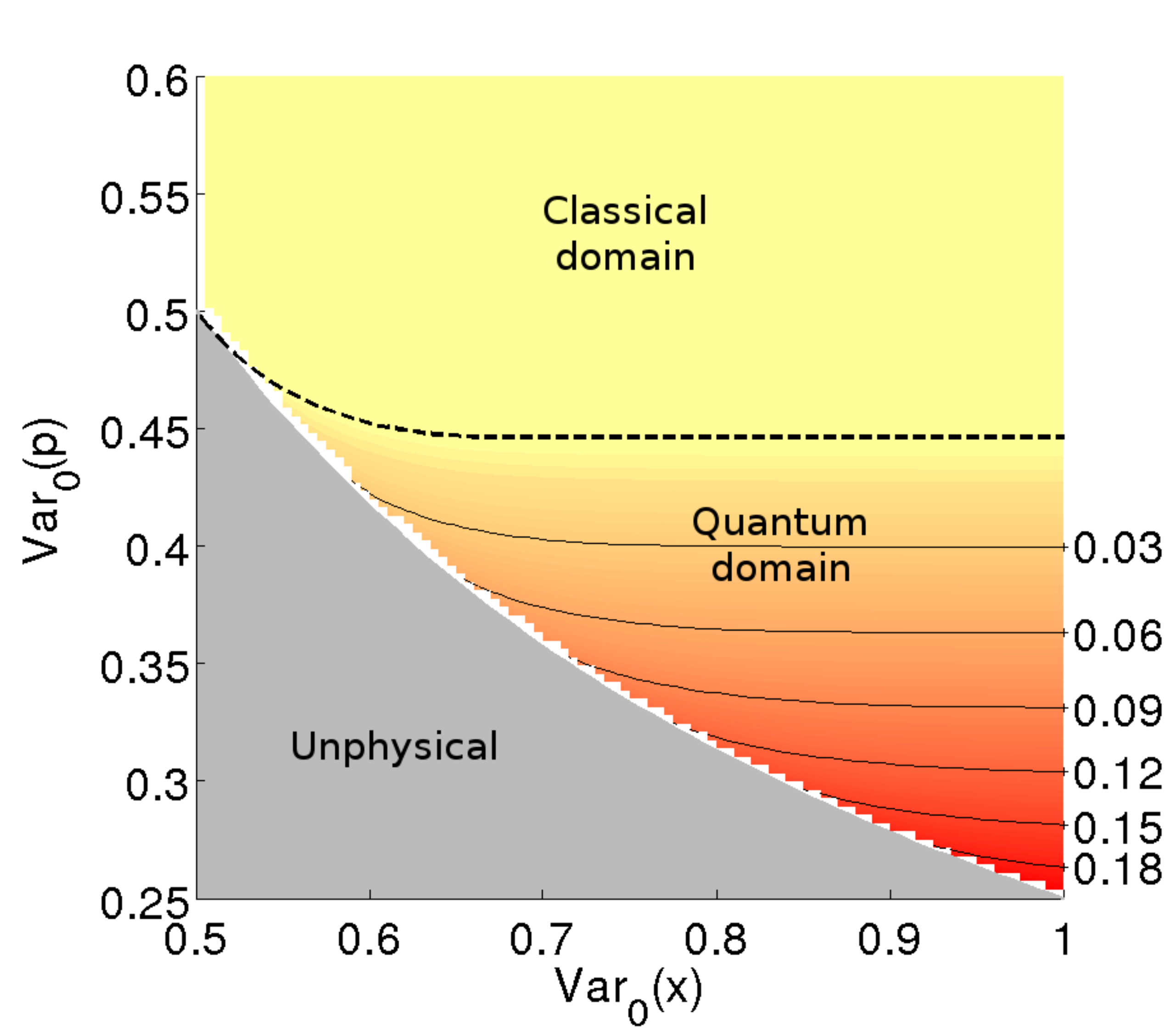}
	\caption{(Color online) Entanglement quantification for squeezed/antisqueezed state benchmarks, using initial squeezing magnitude $r=0.35$ and cutoff $N=20$ (cf. Fig. 1 of \cite{haseler09a}). The bottom left region is forbidden by the uncertainty relation. The quantum domain lies between the forbidden region and the upper dotted line (found with EVM method). The quantitative domain, shown using level curves of the minimal negativity, is very faithful in this parameter regime; the zero contour and the quantum domain boundary coincide within the numerical resolution. For much larger values of $\text{Var}_0(\hat{x})$, the two domains begin to diverge due to reduced numerical accuracy. The small `sawtooth' gap near the unphysical boundary is due to the finite resolution of sampled points.}
	\label{fig:squeezed}
\end{figure}

Another important class of entanglement-based benchmarks uses squeezed and antisqueezed vacua as test states. To simplify the display of results, we restrict ourselves to devices whose operation is phase independent, so that the changes in the long/short quadratures are the same for both test states, regardless of their orientation \cite{haseler09a}. In this case, the relevant parameters for the quantum domain are simply the initial squeezing magnitude $r$ and the output variances of one test state, $\text{Var}_0(\hat{x})$ and $\text{Var}_0(\hat{p})$. Example results for this benchmark type are shown in Fig. \ref{fig:squeezed}, for the parameters $r=0.35$ and $N=20$. We see that the quantitative domain is very faithful with the quantum domain in the given parameter region. For much larger values of the variance than shown in Fig. \ref{fig:squeezed}, the entanglement bounds eventually decay. Again, this is due to the chosen cutoff $N=20$ not being large enough to counter higher noise values.

\begin{figure}
	\includegraphics[width= 1.0 \columnwidth]{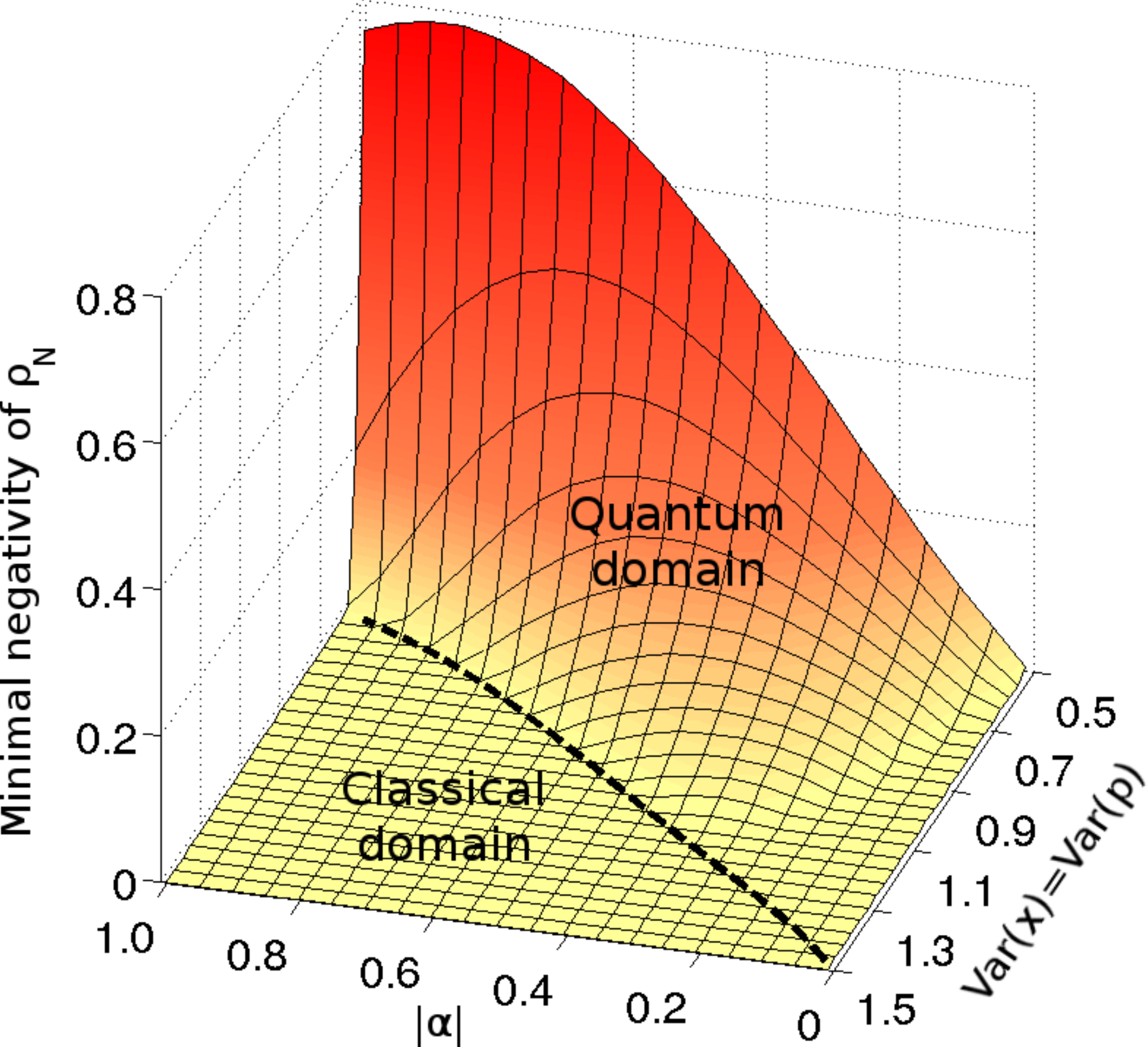}
	\caption{(Color online) Entanglement quantification for the three coherent state benchmark, with no loss ($T=1.0$) and cutoff $N=15$ (cf. Fig. 3 of \cite{haseler10a}). Variance values below the dashed black line are in the quantum domain (note again that the axes are flipped to aid visualization of the results). Although the quantitative domain is larger than the case with two coherent states, it does not cover the entire quantum domain due to the smaller cutoff and higher energies involved.}
	\label{fig:threestates}
\end{figure}

Finally, the last example of entanglement-based benchmarking we will consider is for three coherent states distributed symmetrically on a ring. Benchmarks with multiple coherent states are known to lead to larger quantum domains than those with just two coherent states \cite{haseler10a}. The magnitude $|\alpha|$ of the coherent states is used as an input parameter, while loss and excess noise paramatrize the output. The results of applying our quantification procedure to this scenario are shown in Fig. \ref{fig:threestates}, using $N=15$ (this is lower than in the previous cases because the extra dimension on subsystem A also needs to be accomodated within our available computation resources). We find that the quantitative domain is larger than for two coherent states. However, there is a small but noticable gap between the quantitative domain and the known quantum domain. 

\begin{figure}
	\includegraphics[width= 1.0 \columnwidth]{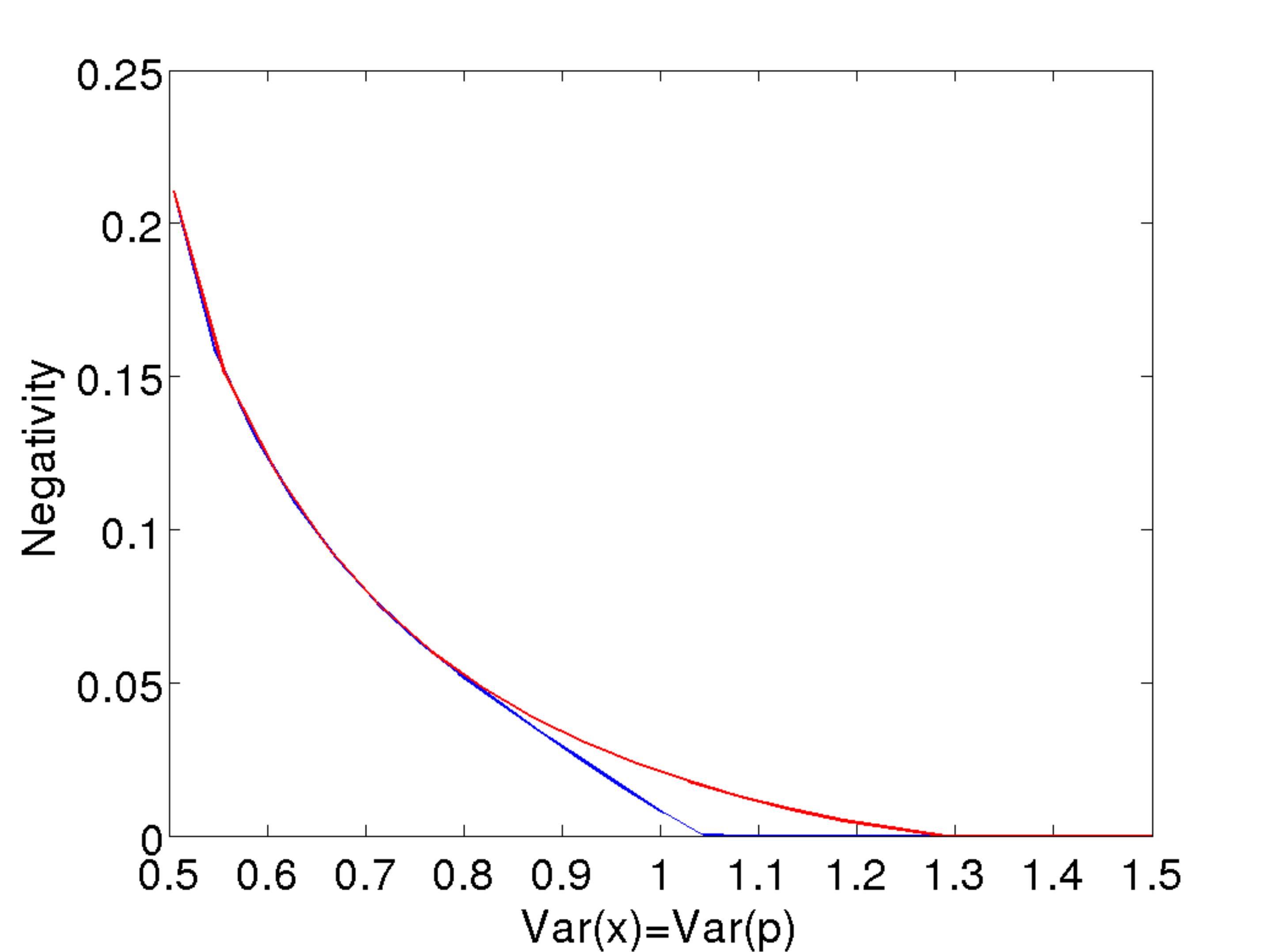}
	\caption{(Color online) Upper and lower bounding the minimal entanglement. The lower curve is a slice through the surface in Fig. \ref{fig:threestates} at $|\alpha|=0.2$ and the upper curve corresponds to results of the hybrid optimization (\ref{eq:hybridopt}). We see that the minimal entanglement must be quite small after $Var(x)\approx1$. Higher values of the cutoff $N$ would be needed to have small enough error bounds to detect this entanglement.}
	\label{fig:hybridopt}
\end{figure}

To explore the strength of our scheme, we consider the following hybrid of the optimizations (\ref{eq:minent}) and (\ref{eq:minentproj}): 
\begin{align}\label{eq:hybridopt}
	 \mathcal{H}(\rho_{AB})=\begin{cases}
	    	\displaystyle\min_{\sigma_{N}\geq 0} &\mathcal{E}(\sigma_{N})\\
	  s.t.~ &\tr(\sigma_{N}\ketbra{i}{i}\otimes\hat{M_k})=\tr(\rho_B^i\hat{M_k})\\
	  &\tr(\sigma_{N}\ketbra{i}{j}\otimes\iden_B)=\bra{j}\rho_A\ket{i}
	 \end{cases},
\end{align}
where, as above, the operators $\hat{M}_k$ are taken from the set $\{\iden,\hat{a},\hat{d}\}$. In this optimization problem, we search through states in the projected subspace (for some fixed value of $N$), but we force the expectation values to be exactly those of the infinite-dimensional state $\rho_{AB}$. The result of this optimization will necessarily be an upper bound on the minimal entanglement of $\rho_{AB}$. In Fig. \ref{fig:hybridopt}, we compare the upper bound found by this hybrid optimization (using $N=15$) with the results shown in Fig. \ref{fig:threestates} by considering a section at amplitude $|\alpha|=0.2$. We see that our result is fairly tight to the upper bound for most variance values. We also see that there is very little negativity to be found (less than 0.03) for $Var(\hat{x})\geq1$. The mismatch between the quantitative domain and the quantum domain again seems to be due to the combination of low entanglement and weaker constraints at high noise values. If desired, the quantitative domain could likely be extended by employing more computational resources.

\section{Conclusion}\label{sec:conclusion}
We have outlined a new method for quantitatively extending entanglement-based benchmarks. The method provides rigorous lower bounds on the entanglement content of test states after partial action by an optical device. The method, which requires no additional information, leads to very good agreement with many of the known quantum domains, providing a huge improvement over the quantification scheme of \cite{killoran10b} (though we should note that the results presented here do not supersede those of \cite{killoran10b}, which appears to be a better choice for devices with very low levels of imperfection). Most importantly, the method presented here enables us, for the first time, to quantitatively characterize the quantumness of realistic optical devices. 

\acknowledgments{We would like to thank M. Piani for useful discussions. We acknowledge support from the NSERC Innovation Platform Quantum Works. As well, N.K. acknowledges support from the Ontario Graduate Scholarship program.}

\appendix
\section{Derivation of first order constraint}\label{app:firstmomentbounds}
Here, we detail how to arrive at the bounds given in Eqs. (\ref{eq:abound}-\ref{eq:epsilon}). Let $\tau$ be a bounded positive-semidefinite operator and let $\tau_N$ be the projection of $\tau$ onto the subspace given by the first $N+1$ Fock states, $\tau_N = \sum_{k=0}^N\bra{k}\tau\ket{k}$. We seek to bound the magnitude of the difference
\begin{align}
	\tr(\tau\hat{a})-\tr(\tau_N\hat{a})   = & \sum_{k=0}^\infty  \bra{k}(\tau-\tau_N)\hat{a}\ket{k}\notag\\
	 = & \sum_{k=0}^\infty\sqrt{k}\bra{k}(\tau-\tau_N)\ket{k-1}.
\end{align}
For simplicity, denote general matrix elements by $\bra{m}\tau\ket{n}=:\tau_{m,n}$, and diagonal entries by $\tau_{m,m}=:\tau_m$. Then, from the triangle inequality, 
\begin{equation}
	\left| \tr(\tau\hat{a})-\tr(\tau_N\hat{a}) \right| \leq  \sum_{k=N+1}^\infty \sqrt{k}\left|\tau_{k,k-1}\right|
\end{equation}
From the positivity of $\tau$, we must have
\begin{equation}
	\tau_m\tau_n\geq\left|\tau_{m,n}\right|^2.
\end{equation}
Therefore,
\begin{equation}
	\left| \tr(\tau\hat{a})-\tr(\tau_N\hat{a}) \right| \leq \sum_{k=N+1}^\infty\sqrt{k}\sqrt{\tau_k}\sqrt{\tau_{k-1}}
\end{equation}
Using the Cauchy-Schwarz inequality on this sum, we find
\begin{equation}
	\left| \tr(\tau\hat{a})-\tr(\tau_N\hat{a}) \right| \leq \sqrt{\left( \sum_{k=N+1}^\infty k\tau_k \right)\left( \sum_{k=N+1}^\infty\tau_{k-1} \right)}.
\end{equation}
The first bracketed term is the difference in mean photon number between $\tau$ and $\tau_N$, while the second is the difference in trace between $\tau$ and its truncation after the level $N-1$. Finally, we arrive at the desired bound:
\begin{equation}
	\left| \tr(\tau\hat{a})-\tr(\tau_N\hat{a}) \right| \leq \sqrt{(\nbar-\nbar_N)(\tr(\tau)-\tr(\tau_{N-1}))}.
\end{equation}

\section{Derivation of second order constraint}\label{app:secondmomentbounds}
In this appendix, we outline how to arrive at the bound on the difference operator found in Eqs. (\ref{eq:differencedifference}-\ref{eq:delta}). We begin with
\begin{align}
	\tr(\tau \hat{a}^{\dagger 2})-\tr(\tau_N \hat{a}^{\dagger 2}) =& \sum_{k=0}^{\infty}\bra{k}(\tau-\tau_N)\hat{a}^{\dagger 2}\ket{k}\notag\\
	 = & \sum_{k=N-1}^{\infty}\sqrt{k+1}\sqrt{k+2}\tau_{k,k+2}.
\end{align}
Similarly, 
\begin{equation}
	\tr(\tau \hat{a}^2)-\tr(\tau_N \hat{a}^2) =\sum_{k=N-1}^{\infty}\sqrt{k+1}\sqrt{k+2}\tau_{k,k+2}^*.
\end{equation}
Therefore, with $\hat{d}=\hat{a}^{\dagger 2} + \hat{a}^2$,
\begin{align}
	\left| \tr(\tau \hat{d})-\tr(\tau_N \hat{d})\right|  = & \left| 2\sum_{k=N-1}^{\infty}\sqrt{k+1}\sqrt{k+2}\text{Re}[\tau_{k,k+2}] \right|\notag\\
	 \leq & ~2\sum_{k=N-1}^{\infty}\sqrt{k+1}\sqrt{k+2}\left|\tau_{k,k+2} \right|.
\end{align}
As in Appendix \ref{app:firstmomentbounds}, we can use the positivity of $\tau$ and the Cauchy-Schwarz inequality to get
\begin{align}
	 &\left| \tr(\tau \hat{d})-\tr(\tau_N \hat{d})\right| \notag\\
	 &\leq  2\sqrt{\left(\displaystyle\sum_{k=N-1}^{\infty}(k+2)\tau_{k+2}\right)\left(\displaystyle\sum_{k=N-1}^{\infty}(k+1)\tau_k\right)}.
\end{align}
Finally, by rewriting the right hand side as
\begin{equation}
	2\sqrt{\left(\nbar-\nbar_N\right)\left[ (\nbar-\nbar_{N-2}) + (\tr(\tau)-\tr(\tau_{N-2})) \right]}=\delta,
\end{equation}
we arrive at the desired bound.

\end{document}